\documentclass[showpacs,preprintnumbers,superscriptaddress,amsmath,amssymb,latexsym,twocolumn]{revtex4}
\usepackage{graphicx}
\usepackage[latin1]{inputenc}
\usepackage[english]{babel}

\def\beq{\begin{equation}}
\def\eeq{\end{equation}}

\begin{document}

\title{Superradiance from BEC vortices: a numerical study}

\author{F. Federici}
\email{fr.federici@sns.it}
\affiliation{NEST-INFM and Classe di Scienze, Scuola Normale
  Superiore, Piazza dei Cavalieri 7,  56126 Pisa, Italy}
\author{C. Cherubini}
\affiliation{Facolta' di Ingegneria, Universita' Campus Biomedico di Roma,
Via Longoni 83, 00155 Roma, Italy}
\author{S. Succi}
\affiliation{Istituto per le Applicazioni del Calcolo, CNR, Viale del
  Policlinico 137, 00161 Roma, Italy}
\author{M.P. Tosi}
\affiliation{NEST-INFM and Classe di Scienze, Scuola Normale
  Superiore, Piazza dei Cavalieri 7,  56126 Pisa, Italy}

\begin{abstract}

The scattering of sound wave perturbations 
from vortex excitations of Bose-Einstein condensates
(BEC) is investigated by numerical integration of the associated Klein-Gordon
equation. It is found that, at sufficiently high angular speeds, 
sound wave-packets can extract a sizeable fraction of  
the vortex energy through a mechanism of superradiant scattering.  
It is conjectured that this superradiant regime may be detectable 
 in BEC experiments. 
\end{abstract} 

\pacs{04.70.-s, 03.75.Lm} 

\maketitle

Recent years have witnessed a growing interest in pursuing 
analogue models of gravitational physics in condensed 
matter systems. 
The rationale for such models traces back to a seminal observation 
by Unruh \cite{ref41}, who noted a close analogy between 
sound wave propagation in an  inhomogeneous background
flow and field propagation in curved 
space-time. The analogy goes on by observing that, much like 
superfluid hydrodynamics is a large-scale effective theory 
of microscopic superfluids, field theory on a curved space-time 
might also be regarded as a large-scale limit of a possible microscopic 
formulation of quantum gravity. The crucial point is that, whereas
microscopic theories of quantum gravity are still largely a matter of 
speculation, the microscopic theory of superfluids is well developed. 
It can thus be hoped that the wide body of knowledge 
available for the latter can be brought to the benefit of the 
former \cite{zoller}.  
For instance, assessing the mechanisms of sound radiation 
from `terrestrial black holes' beyond the hydrodynamic picture 
may in principle offer new insights into the microscopic 
origin of cosmic black hole radiance, the Hawking effect, and 
other cosmological phenomena. 

A key step along this long-term program  
is the study of scattering and radiance 
phenomena from black holes whose background space-time can be 
associated with fluid excitations such as vortices. 
A model of fluid flow which seems particularly well 
suited to pursue the 'analogue gravity' program is the 
so-called draining-bathtub  geometry \cite{Visser},  
namely a three-dimensional flow with a sink (vortex)  
at the origin. The flow field induced by 
the vortex is associated with an acoustic metric with 
two crucial ingredients of black hole physics:  
an event horizon and an ergosphere.  
The former is a spatial surface which allows only one-way propagation 
of physical signals
(from the outside into the vortex), while 
the latter is a region from which part of the 
the vortex energy can be extracted via the mechanism of 
superradiance. 

Such a phenomenon was first studied by Zel'dovich
\cite{zel} with regard to the generation of waves by a rotating body
and was then
analysed as stimulated emission in black-hole radiance 
\cite{Star, Dewitt,Wald}. Superresonance is an
acoustic-wave version of the Penrose \cite{penrose} process, 
whereby a plane-wave
solution of a scalar massless field in the black hole background is
scattered from the ergosphere with an amplification at the expenses 
of the rotational energy of the black hole.
Such process has been shown to occur in a certain class of analogue
(2+1)-dimensional rotating black holes \cite{ref41}. Later studies
\cite{Basak1,Berti} have discussed the frequency dependence of the
amplification factor in superresonant scattering of acoustic
perturbations from a rotating acoustic black hole by deriving the
reflection coefficient as a
function of the frequency $\sigma_0$ of the incoming monochromatic wave. It is
found that in the range $0<\sigma_0 <m\Omega$ the reflection
coefficient is greater than unity, with $m$ being the azimuthal wavenumber and
$\Omega$ the angular frequency of the acoustic horizon.
  
The main purpose of this paper is to present a quantitative 
investigation of superradiant scattering from sonic holes 
associated with BEC-like vortex configurations. 
Although superradiant scattering from hydrodynamic vortices 
has been discussed in the recent literature \cite{SAVAGE,BASAK3}, 
we believe that this is the first quantitative assessment of such 
a phenomenon under specific BEC-like conditions.

In the limit of zero temperature, gaseous Bose-Einstein condensates  
are well described by the Gross-Pitaevskii equation (GPE)
\begin{equation}
i \hbar \partial_t \Phi =\left( -\frac{\hbar^2}{2M} \nabla^2  
+ V_{\rm ext}+ \frac{4\pi\hbar^2a_{\rm s}}{M}|\Phi|^2 \right)\Phi, 
\label{GPE}
\end{equation}
where $\Phi(\vec r,t)$ is the wavefunction of the condensate
normalised to the total number of bosons $N$,
 $a_s$ being the s-wave
scattering length  and $M$ the mass of the atoms. 
If we now use the Madelung representation $\Phi(\vec
r,t)=\sqrt{\rho(\vec r,t)}e^{i M\theta(\vec r,t)/\hbar}$ 
\cite{mad} in Eq. (\ref{GPE}), where
 $\rho(\vec r,t)=|\Phi(\vec r,t)|^2$ is the condensate density, 
the GPE equation takes a hydrodynamic form: the imaginary part is a continuity
equation for an irrotational fluid flow of velocity 
$\vec{v}(\vec r,t)=\nabla \theta(\vec r,t)$ and density $\rho(\vec r,t)$,
and the real part is a Hamilton-Jacobi equation whose gradient leads
to the Euler equation. As is well known, the GPE
is equivalent to irrotational inviscid hydrodynamics \cite{stringari}.

Low-frequency perturbations around the stationary state 
 are essentially sound waves (zero sound) and
obey the
Bogoliubov set of differential equations
for the density perturbation $\rho^{(1)}$ and the phase perturbation
$\theta^{(1)}$ in terms of the local speed of sound 
$c(\vec r)=\sqrt{4\pi \hbar^2 a_s\rho(\vec r)/M^2}$. These equations,
within the limit of validity of the hydrodynamic approximation
 can be reduced to a single second-order equation for the
phase perturbation \cite{zoller}. This
differential equation for $\theta^{(1)}\equiv\Psi$ has the form of a 
relativistic Klein-Gordon
equation
$\partial_{\mu}\left(\sqrt{-g}g^{\mu\nu}\partial_{\nu}\Psi\right)=0$,
with $g={\rm det}\, g_{\mu\nu}$ in a curved space-time whose
metric $g_{\mu\nu}$ is determined by the local speed of
sound $c$ and the background stationary velocity $\vec v$. 

It should be noted that the linearization suppresses the quantum
 nature of the GPE so that, within the linear perturbation theory, 
the circulation of vortices is  not quantised as in BEC systems. 
The calculations reported below are aimed at examining  what fraction of
the energy can be extracted through superradiant scattering of a sound wave
from a vortex described by BEC-like parameters.
As discussed further below,  the full nonlinear GPE will have to be
 used for a quantitative assessment of the extraction of energy quanta 
from  BEC vortices.

For a single vortex with a drain at $r=0$
and angular velocity $\Omega$ in the draining-bathtub model,
 the velocity field of the flow is 
\begin{equation}
\vec v=\nabla \theta(r,\phi)=\left(-ca \hat r+\Omega a^2\hat\phi\right)/r\,. 
\label{VEL} 
\end{equation} 
where $\hat r$ and $\hat \phi$ denote unit vectors in polar
coordinates, $a$ is the radius of the event horizon, and 
the background density $\rho_0$ of the fluid and the speed of sound $c$
are  taken as constant throughout the flow.
The acoustic metric associated with this configuration is
\begin{eqnarray}
\label{metrica}
ds^2=&&-\left(c^2- ((a^2 c^2+ a^4\Omega^2)/r^2)\right)dt^2+ 
(2c a/r)\,dt\,dr\nonumber \\
&&-2\,\Omega a^2 dt\, d\phi+dr^2+r^2d\phi^2+dz^2.
\end{eqnarray} 
It is readily checked that this metric has  
an ergosphere whose radius is $r_{erg}=a\sqrt{1+\Omega^2 a^2/c^2}$.
The growth of the ergosphere with increasing $\Omega$ allows an
increasing extraction of energy from the vortex in superradiance
conditions. 

Linear perturbations of the velocity  
potential $\Psi$ satisfy the  
massless Klein-Gordon scalar wave equation  
on this background, i.e. 
\begin{eqnarray} 
\label{SCALEQ} 
&&\left[-\frac{1}{c^2}\frac{\partial^2}{\partial t^2}+\frac{2 a}{cr}\frac{\partial^2}{\partial t\partial r}-\frac{2 a^2\Omega}{c^2 r^2}\frac{\partial^2}{\partial t \partial \phi}+\left(1-\frac{ a^2}{r^2}\right)\frac{\partial^2}{\partial r^2}+\right.\nonumber\\ 
&&\left.+\frac{2 a^3\Omega}{cr^3}\frac{\partial^2}{\partial r\partial \phi}+ 
{\frac {{c}^{2}{r}^{2}-{ 
      a}^{4}{\Omega}^{2}}{{c}^{2}{r}^{4}}}\frac{\partial^2}{\partial 
  \phi^2}+\frac{\partial^2}{\partial z^2}+\right. \\ 
&&\left. +{\frac {r^2+{ a}^{2}}{{r}^{3}}}\frac{\partial}{\partial r}-\frac{2 
  a^3\Omega}{cr^4}\frac{\partial}{\partial \phi} 
\right]\Psi=0\,.\nonumber 
\end{eqnarray} 
Equation (\ref{SCALEQ}) is solved by means of numerical methods 
developed  in \cite{TEUKO} for the integration of massless
scalar-field  
perturbations on a rotating Kerr black-hole background. 
To this purpose, Eq. (\ref{SCALEQ})
is  conveniently recast into a system of first-order 
(strongly) hyperbolic equations through the definition 
of two conjugate fields $\Xi_i=\partial\Psi/\partial x^{i}$ and 
$\Pi=-(1/\alpha)\left(\partial\Psi/\partial t -\beta^{i}\Xi_{i}\right)$, 
where $\beta^i=({ ac}{r^{-1}},-{ a^2\Omega}{r^{-2}},0)$ and 
$\alpha =c$ are the space and time shifts of the acoustic metric. 
By setting 
$\Psi=\psi_1(r,t)e^{im\phi}e^{ikz}$, $\Pi=\pi_1(r,t)e^{im\phi}e^{ikz}$, 
$\Xi_1=\xi_1(r,t)e^{im\phi}e^{ikz}$,  $\Xi_2=im\Psi$, and 
$\Xi_3=ik\Psi$, 
where $(k,m)$ are the axial and azimuthal wavenumbers, 
the hyperbolic system reads as follows: 
\begin{eqnarray} 
\label{teuk}
&&\partial_{t}\pi_1+c\partial_{r}\left(\xi_{1}- a\pi_{1}/r\right)=
\left(ac-im a^2\Omega\right)\pi_1/r^2+\nonumber\\
&&+c\left(k^2+m^2/r^2\right)\psi_{1} -c\xi_{1}/r \nonumber\\
&&\partial_{t}\psi_{1} -c\partial_{r}\left(a\psi_{1}/r\right)=
\left(ac-im a^2\Omega\right)\psi_{1}/r^2-c\pi_{1} \\
&&\partial_{t}\xi_{1}+c\partial_{r}\left(\pi_{1}-a\xi_{1}/r\right)=
2im a^2 \Omega\psi_{1}/r^3-ima^2\Omega\xi_{1}/r^2\nonumber\, 
\end{eqnarray} 
The set of Eqs. (\ref{teuk}) is augmented with the constraint  
$\vert C\vert\equiv\vert\partial_r\psi_1-\xi_1\vert=0$, 
which is used to monitor the quality of the numerical 
results. One-way inward 
propagation from the horizon is accounted for by an ingoing-radiation  
boundary condition, imposed through an excision technique.  
Details of the numerical procedure will be given in a forthcoming 
publication \cite{CFST}. 
 
Following the standard prescription  
for scattering processes in Kerr black holes \cite{Laguna},  
the initial condition is chosen as 
a Gaussian pulse centered at $r=r_0$ and modulated by a monochromatic
wave,
\begin{equation} 
\psi_1(r,0)=A\exp[{-{(r-r_0+ct)^2}/{b^2}-i\sigma(r-r_0+ct)/c}]\vert_{t=0}.
\end{equation} 
The corresponding power spectrum is a Gaussian distribution 
$P(\omega)=P_{{\rm max}}\exp{\left[-(\omega-\sigma)^2b^2/4c^2\right]}$, 
centered at frequency $\sigma$ with spectral width $1/b$. The
superradiant regime is $0<\sigma /\Omega<m$ for $m\geqslant 1$. 
  
As already remarked, the main purpose of our calculations is to assess
the amount of energy that 
can be extracted from a BEC-like sonic hole as a function of its 
angular velocity $\Omega$. 
We use as reference a set of parameters relevant to a 
BEC of $^{87}$Rb atoms \cite{Dalibard}, with  
vortex core radius $a\simeq 0.2$ $\mu$m  and angular speed 
$\Omega \simeq 18$ KHz.  
It is interesting to note that the inverse transversal time of the BEC
vortex, $c/a \sim 15$ KHz, is very close to
the corresponding value for a cosmic black hole of radius 
$a \sim 10$ km.
Such a quantitative match stems from the very low speed of sound
in BEC's, of the order of a few mm/s.       
 
In the following we take $c=\hbar/(\sqrt{2}M\xi)$ and $a=\xi$, where 
$\xi=\left(8\pi \rho a_{\rm s}\right)^{-1/2}$ is the healing length.
The integration of Eqs. (\ref{teuk}) is performed in the space-time 
domain $r\in[0,\, 150]$ and 
$t\in[0,\, 150]$, in units of $a=1$ and $a/c=1$ in space and
time. 
The angular frequency is analysed in the range 
$0.14 < \Omega a/c < 14$, corresponding to a 
frequency range  $1.8 < \Omega < 18$ KHz and a 
density range $5\times 10^{13} < \rho_0 < 5\times 10^{14}$ $cm^{-3}$. 
The initial Gaussian pulse is centered at $r_0=50a$ and $\sigma=0.5\Omega$, 
with amplitude $A=0.3c$ and variance $b=10a$. 
We perform our study for $m=1$ and $k=0.02/a$, corresponding to a condensate
 axial extent $H=0.9$ mm.    
Violations of the constraint $C(t)=0$ are monitored over the
entire space domain and are  found 
to be consistently below $10^{-6}$ for all sets  
of parameters under investigation.  
 
\begin{figure}
\centering 
\includegraphics[width=8.5cm]{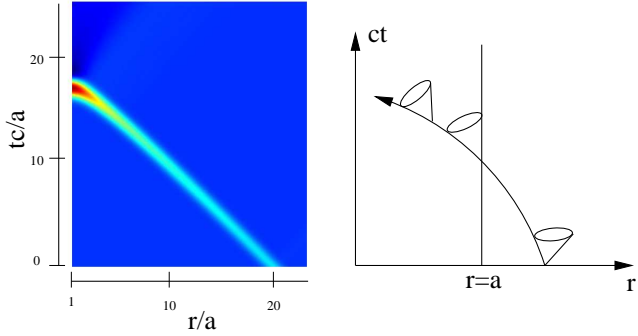} 
\caption{A density map of the real part of $\psi_1$ 
in the $r-t$ plane (for the sake of clarity, the case of a
quasi-localised wavepacket at $m=0$ is being shown).
Due to the curved background, the wavepacket trajectory bends towards
the sonic horizon at $r=a$. The light cone becomes parallel to 
the $r=a$ axis, since no signal can escape
from the horizon.} 
\label{Fig1} 
\end{figure}

Fig.\ref{Fig1} shows a density map of the real part of $\psi_1$  
in the range $r/a\in[1,\, 20]$ and $tc/a\in[0,\, 10]$.  
The initial Gaussian pulse moves towards the  
vortex horizon placed at $r=a$ and its trajectory 
is bent by the potential outside the horizon. 
The bending of the trajectory, with light-cones heading towards 
the horizon, is consistent with similar findings in numerical 
relativity \cite{Wald,text3} as shown in the diagram in Fig.\ref{Fig1}.  
  
\begin{figure} 
\centering 
\includegraphics[width=8.5cm]{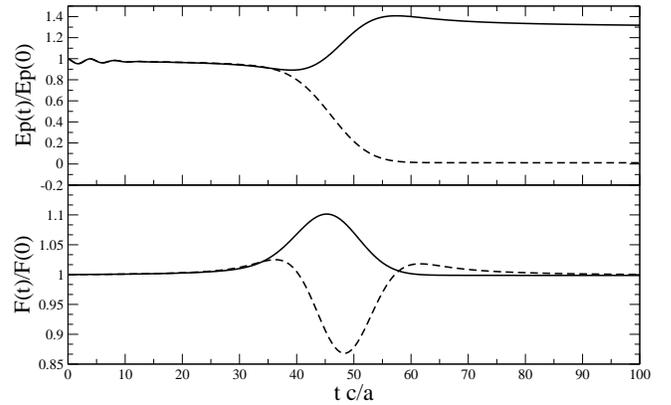} 
\caption{Time evolution of the energy gain (top) and corresponding
  fluxes (bottom) of the wavepacket 
for $\sigma=0.7c/a$ and $\Omega=1.4c/a$. The solid lines correspond to the
superradiant  case ($m=1$) and the dashed lines to the non-superradiant
one ($m=0$).}
\label{Fig2} 
\end{figure}

In Fig.\ref{Fig2} we show a typical time evolution of the energy of 
wavepacket 
$E_{\rm p}(t)=(\rho_0 M/2)\int_0^{2\pi} \,{\rm d}\phi \int_{0}^{H}\, {\rm d}z
\int_{a}^{143a} v_1^2 r\,{\rm d}r$ with $v_1=\nabla \theta^{(1)}$ (top
curves),  normalized to its initial value $E_p(0)$, as well as the
(independently calculated) rate $F(t)$ of change of the energy  (bottom
 curves), normalized to its initial value $F(0)$, for $\sigma=0.7c/a$ and 
$\Omega=1.4c/a$, within the superradiant regime ($m=1$, solid lines) and 
outside it ($m=0$, dashed lines).  $F(t)$
  includes the net flux across the surfaces at $r=a$ and $r=143a$
  as well as a term due to the bulk compressibility,
\begin{eqnarray}
\label{lux}
F(t)&&=\frac{{\rm d}E_{\rm p}(t)}{{\rm d}t}=\int\vec v_1\cdot\frac{{\rm d}\vec
  v_1}{{\rm d}t}\, {\rm d}V =\\
&&= -\frac12\left[\int v_1^2\vec v\cdot\hat n \,{\rm d}S-\int v_1^2
  \nabla\cdot\vec v_1\,{\rm d}V\right].\nonumber
\end{eqnarray}
In the non-superradiant case, the energy of the scattered 
wavepacket goes asymptotically to zero, indicating 
that all the energy of the impinging wavepacket is lost  
to the vortex sink.   
In the superradiant case instead,
the energy of the back-scattered wavepacket exceeds its initial value, 
indicating  extraction of  energy from the  
ergosphere at the expense of the rotational energy of the vortex. 
Consistently with this picture, the energy flux for the superradiant
(non-superradiant) case lies 
above (below) its background value during the scattering event, 
approximately in the range $35<tc/a<55$. 
The energy gained via 
superradiance is by no means small, as it is seen to 
exceed in this case twenty percent of the initial value $E_p(0)$. 

\begin{figure}
\centering 
\includegraphics[width=8.5cm]{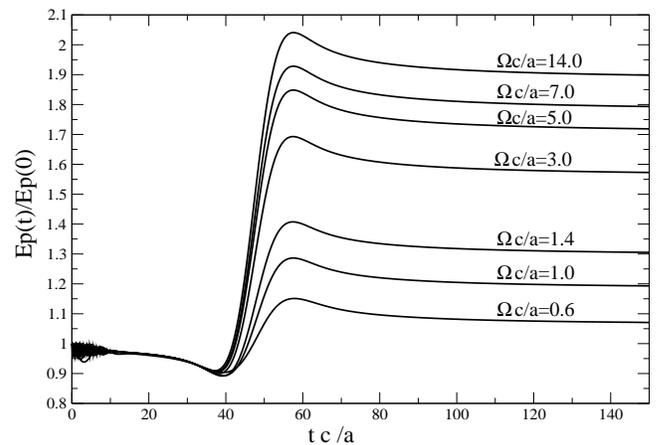} 
\caption{Time evolution of the energy gain for different values of $\Omega$
  all in the superradiant regime. 
Here $m=1$, $b=10a$, $r_0=50a$, and $\sigma=0.5\Omega$.} 
\label{Fig3} 
\end{figure}

It is now of great interest to examine the dependence 
of the superradiant energy gain on the angular speed of the vortex, 
so as to possibly identify an optimal value  
at which such energy gain can be maximised.  
In Fig.\ref{Fig3} we show the time evolution of the energy gain 
$E_p(t)/E_p(0)$ for a series of values of $\Omega$ in the superradiant range 
$\Omega a/c\in [0.6,\,14]$, as can be experimentally achieved by varying
the density of the condensate.  
A sharp increase of the energy  
gain is observed in the region $\Omega > c/a$. 
This is plausible, since $\Omega > c/a$ marks a transition  
from the regime where the radius of the ergosphere  
remains within a factor two of the sonic 
horizon, to the regime where it grows linearly  
with $\Omega$, thereby creating a sizeable  
ergospheric shell where energy can be extracted from. 

\begin{figure}
\centering 
\includegraphics[width=8.5cm]{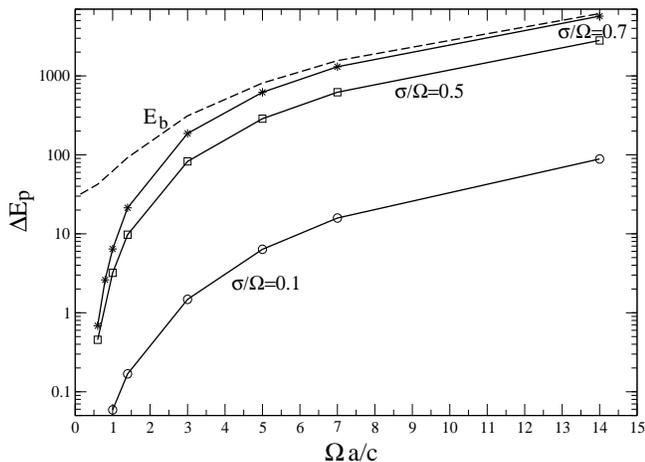} 
\caption{$\Delta E_{\rm p}$ is plotted in units of $\rho_0 Mc^2a^2H/2$ on 
a logarithmic scale 
 as a function of $\Omega a/c$ for
  three values of $\sigma/\Omega$ (solid lines)  and
  compared with the background vortex energy $E_{\rm b}$ on the same
  scale and in the same units (dashed line).} 
\label{Fig4} 
\end{figure}

Since BEC vortices are quantized, one is naturally led  
to ask whether the wavepacket may annihilate 
the vortex by extracting all of its energy in a  much more demanding
 'break-even' condition, that is 
$E_{\rm p}(\infty)-E_{\rm p}(0)=E_{\rm b}$ where $E_{\rm b}$ denotes the  
energy of the background vortex. 
In Fig. \ref{Fig4} we show the background energy $E_{\rm b}$ (dashed
 line) and the total energy gain
$\Delta E_{\rm p}=(E_{\rm p}(\infty)-E_{\rm p}(0))$  for three
values of $\sigma/\Omega$ in the superradiant range(solid lines), 
as  functions of $\Omega$.
In the perturbative regime (for $\Omega\leq 3c/a$, say) the
efficiency of energy extraction from the vortex grows much
faster
with $\Omega$ than the quadratic increase of the background
energy. This is especially true at large values of the ratio
$\sigma/\Omega$.
Although substantial values of $\Delta E_{\rm p}/E_{\rm b}$ are - by
definition - beyond the scope of 
the perturbative Klein-Gordon  
description used throughout this work, it appears that nonlinear
effects may primarily determine the way in which the energy extraction
behaves as it becomes comparable to the background energy.
The possibility that substantial superradiance
efficiencies, as they emerge
from the Klein-Gordon analysis, may persist even in  
the non-perturbative quantum regime described by the   
GPE cannot be ruled out. 
Even though the condition $\Delta E_{\rm p}=E_{\rm b}$ may
remain out of reach for a single wavepacket, one may
still conjecture that a train of wavepackets 
could reach the goal. 
It would be interesting to test this
conjecture by numerical and experimental means. 

As a further development of the present model, it will also be
interesting to consider a vortex with no drain \cite{SAVAGE,Volovik}
and to apply our analysis to superradiant scattering 
 from a giant vortex \cite{giant,Simula,Tsubota}. Such vortices have
 been found to have up to  60 quanta of
circulation and can therefore be well approximated within the
classical limit.

In summary, numerical simulations of the Klein-Gordon equation for 
sonic perturbations impinging on a BEC-like vortex 
suggest the possibility that, under typical 
conditions of BEC experiments, a significant fraction of the 
vortex energy may be extracted via the mechanism of superradiance.  
It would be very interesting to test the  realisability of such an
exciting scenario both via the numerical solution of the
Gross-Pitaevski equation and by actual experiments on rotating
Bose-Einstein condensates.

\begin{acknowledgments} 
This work was partially supported by an Advanced Research Initiative
of SNS.
\end{acknowledgments}

\end{document}